\documentclass[12pt]{article}
\def\be{\begin{equation}}
\def\ee{\end{equation}}
\def\bea{\begin{eqnarray}}
\def\eea{\end{eqnarray}}
\usepackage{graphicx}

\catcode`\@=11
\def\lsim{\mathrel{\mathpalette\@versim<}}
\def\gsim{\mathrel{\mathpalette\@versim>}}
\def\@versim#1#2{\vcenter{\offinterlineskip
\ialign{$\m@th#1\hfil##\hfil$\crcr#2\crcr\sim\crcr } }}
\catcode`\@=12

\catcode`@=12
\begin{document}

\thispagestyle{empty}
\begin{flushright}
UCRHEP-T516\\
March 2012\
\end{flushright}
\vspace{0.3in}
\begin{center}
{\Large \bf Dark-Matter Fermion from Left-Right Symmetry\\}
\vspace{1.0in}
{\bf Ernest Ma\\}
\vspace{0.2in}
{\sl Department of Physics and Astronomy, University of California,\\ 
Riverside, California 92521, USA\\}
\vspace{0.1in}
{\sl Institute for Advanced Study, Hong Kong University of Science and 
Technology,\\ Hong Kong, China\\}
\vspace{0.1in}
{\sl Institute of Advanced Studies, Nanyang Technological University, 
Singapore\\}
\end{center}
\vspace{1.0in}
\begin{abstract}\
In an unconventional realization of left-right symmetry, the particle 
corresponding to the left-handed neutrino $\nu_L$ (with $SU(2)_L$ 
interactions) in the right-handed sector, call it $n_R$ (with $SU(2)_R$ 
interactions), is not its Dirac mass partner, but a different particle 
which may be a dark-matter candidate. In parallel to leptogenesis in the 
$SU(2)_L$ sector, asymmetric production of $n_R$ may occur in the $SU(2)_R$ 
sector.  This mechanism is especially suited for $n_R$ mass of order 1 to 10 
keV, i.e. warm dark matter, which is a possible new paradigm  for 
explaining the structure of the Universe at all scales.
\end{abstract}

\newpage
\baselineskip 24pt

In our present Universe, there is quite a bit of matter, but not much 
antimatter, i.e. we have a matter-antimatter asymmetry, usually called 
the baryon asymmetry.  There is also a lot of dark matter, but its nature 
is unknown.  It is possible that dark matter is related to ordinary 
matter in that it comes partly also from matter carrying baryon number so 
that there is also dark antimatter.  In that case, the mechanism 
which generates the dark matter of the Universe may well be the parallel 
of that which generates the baryon asymmetry. 

It is now a fact of particle physics that neutrinos have mass.  There are 
many theoretical mechanisms~\cite{m09} for this to happen, but the first 
and simplest idea remains that of the canonical seesaw~\cite{seesaw}, where 
a heavy singlet fermion, often referred to as a right-handed neutrino, is 
added to the standard model for each family of leptons.  This implies the 
extra terms given by
\begin{equation}
{\cal L}_\nu = -f (\bar{\nu}_L \phi^0 - \bar{e}_L \phi^+) \nu_R - {1 \over 2} 
M \nu_R^T \nu_R + H.c.,
\end{equation}
which result in a Majorana neutrino mass $m_\nu \simeq - m_D^2/M$, where 
$m_D = f \langle \phi^0 \rangle = fv$.  This very famous simple formula 
also shows that if $M$ is large, $m_\nu$ is small because the Dirac mass 
$m_D$ comes from electroweak symmetry breaking, i.e. $v = 174$ GeV, as 
do all quark and charged lepton masses.  A large $M$ also has a second 
very important consequence.  It allows the decay of $\nu_R$ to 
$e^\pm \phi^\mp$, which may then create a lepton asymmetry in the early 
Universe~\cite{fy86}, and gets converted into the present observed 
baryon (matter-antimatter) asymmetry during the $SU(2)_L$ phase 
transition~\cite{krs85}. 

Consider now the conceptually attractive idea of a left-right extension of 
the standard model.  The existence of $\nu_R$ becomes mandatory and it 
appears as part of the $SU(2)_R$ doublet, i.e. $(\nu,e)_R$.  If $SU(2)_R$ 
is broken at a high scale by a Higgs triplet 
$(\Delta^{++},\Delta^+,\Delta^0)$, then again $\nu_R$ acquires a large 
Majorana mass proportional to $\langle \Delta^0 \rangle$.  The seesaw 
mechanism works as before but it is very difficult to distinguish this 
left-right extension experimentally from the standard model  
at energies far below $M$.

Besides neutrino mass and matter-antimatter asymmetry, the nature and origin 
of dark matter are two equally important issues. Whereas there are myriads of 
theoretical proposals for what dark matter is, recent astrophysical 
observations regarding the structure of the Universe at all scales are 
lending increasing credence~\cite{dvs11} to the notion that it is warm 
and suitably explained by a sterile neutrino~\cite{k09} with mass in the 
range 1 to 13 keV.  In this paper, it is proposed that $\nu_R$ in a 
suitably extended left-right model may in fact behave as a sterile 
neutrino which is absolutely stable in its own right.  This notion is 
fundamentally different from the usual concept of a sterile neutrino which 
mixes with the active neutrinos.  That would make the sterile neutrino 
unstable and not suitable as a dark-matter candidate unless 
the mixing is very tiny.   This mixing also exists in a recent 
left-right extension~\cite{bhl10} where the lightest $\nu_R$ has a  
Majorana mass of order keV.

The origin of the idea that $\nu_R$ does not mix with $\nu_L$ started 
with superstring-inspired $E_6$ models. 
It was realized that with the particle content of the fundamental 
\underline{27} representation, an alternative left-right model is 
possible~\cite{m87}.  More recently, its relevance to dark matter 
in two simple nonsupersymmetric versions~\cite{klm09,klm10} was 
discussed.  In both cases, the $SU(2)_R$ doublet is $(n,e)_R$ where 
$n_R$ is not the Dirac mass partner of $\nu_L$.  It has been termed 
a ``scotino'', i.e. a dark fermion.  In the first model~\cite{klm09}, 
$n_R$ is a Majorana particle of perhaps 150 GeV.  In the second 
model~\cite{klm10}, it is a Dirac particle of perhaps 500 GeV. 
In this paper, a third simple variation is offered such that $n_R$ 
is a Majorana particle of 1 to 10 keV.  As such, it will be shown that 
a $\nu_L-n_R$ correspondence exists in left-right symmetry in all 
their fundamental interactions.  The apparent difference, i.e. $\nu_L$ is 
``active'' and $n_R$ is ``sterile'', only comes from the fact that the 
$SU(2)_R$ breaking scale is much higher than that of $SU(2)_L$.

Let the gauge symmetry $SU(3)_C \times SU(2)_L \times SU(2)_R \times U(1)_{B-L}$ 
be supplemented by a global U(1) symmetry $S'$.  The fermion and scalar 
particle content of this model is listed in Table.~1.
\begin{table}[htb]
\begin{center}
\begin{tabular}{|c|c|c|c|c|c|}
\hline
particles & $SU(3)_C$ & $SU(2)_L$ & $SU(2)_R$ & $(B-L)/2$ & $S'$ \\ 
\hline
$(u,d)_L$ & 3 & 2 & 1 & 1/6 & 0 \\ 
$(u,h)_R$ & 3 & 1 & 2 & 1/6 & --1/2 \\ 
$d_R$ & 3 & 1 & 1 & --1/3 & 0 \\ 
$h_L$ & 3 & 1 & 1 & --1/3 & --1 \\ 
\hline
$(\nu,e)_L$ & 1 & 2 & 1 & --1/2 & 0 \\ 
$(n,e)_R$ & 1 & 1 & 2 & --1/2 & 1/2 \\ 
$\nu_R$ & 1 & 1 & 1 & 0 & 0 \\ 
$n_L$ & 1 & 1 & 1 & 0 & 1 \\
\hline
$(\phi_L^+,\phi_L^0)$ & 1 & 2 & 1 & 1/2 & 0 \\ 
$(\phi_R^+,\phi_R^0)$ & 1 & 1 & 2 & 1/2 & 1/2 \\ 
$\eta$ & 1 & 2 & 2 & 0 & --1/2 \\ 
\hline
\end{tabular}
\caption{Particle content of the fermions and scalars of this unconventional 
left-right model.}
\end{center}
\end{table}
The purpose of the imposed $S'$ symmetry is to distinguish the scalar 
bidoublet $\eta$ from its dual $\tilde{\eta} = \sigma_2 \eta^* \sigma_2$ 
which transforms in the same way as $\eta$ under $SU(3)_C \times SU(2)_L 
\times SU(2)_R \times U(1)_{B-L}$.  Specifically,
\begin{equation}
\eta = \pmatrix{\eta_1^0 & \eta_2^+ \cr \eta_1^- & \eta_2^0}, ~~~~ 
\tilde{\eta} = \pmatrix{\bar{\eta}_2^0 & -\eta_1^+ \cr -\eta_2^- & 
\bar{\eta}_1^0}.
\end{equation}
As a result, $\eta_2^0$ coulples to $\bar{e}_L e_R$, 
$\bar{\eta}_2^0$ coulples to $\bar{u}_L u_R$, 
$\eta_1^0$ coulples to $\bar{\nu}_L n_R$, 
$\bar{\eta}_1^0$ coulples to $\bar{d}_L h_R$, 
$\phi_L^0$ couples to $\bar{d}_L d_R$, 
and $\phi_R^0$ couples sto $\bar{h}_R h_L$. 
As $\phi_R^0$ acquires a large vacuum expectation value, 
$SU(2)_R \times U(1)_{B-L}$ is broken to $U(1)_Y$ and $S'$ is also broken, 
but the combination $S = S' + T_{3R}$ remains unbroken.  To break $SU(2)_L 
\times U(1)_Y$ to $U(1)_Q$, the scalar fields $\phi_L^0$ and $\eta_2^0$ 
(both having $S=0$) acquire vacuum expectation values but not $\eta_1^0$ 
which has $S=-1$.  This means that $n_R$ (which has $S=1$) is not the 
Dirac mass partner of $\nu_L$.

It is easy to see that all the standard-model particles have $S=0$. 
Some of the new particles also have $S=0$, i.e. $Z'$, $\eta_2^+$, $\eta_2^0$, 
$\phi_R^0$, whereas the others have $S=1$, i.e. $n$, $\bar{h}$, $W_R^+$, 
$\phi_R^+$, $\eta_1^+$, $\bar{\eta}_1^0$.  As for baryon number and 
lepton number, $u,d,h$ all have $B=1/3$, and $\nu,e,n$ all have $L=1$. 
Note that $\nu$ gets a Dirac mass from $\phi_L^0$ and $n$ gets a Dirac mass 
from $\phi_R^0$.  Using the canonical seesaw mechanism, the singlets 
$\nu_R$ ($L=1,S=0$) and $n_L$ ($L=1,S=1$) are assumed to have large 
Majorana masses which break $L$ softly to $(-1)^L$ and $S$ softly to 
$(-1)^S$.  Thus both $\nu_L$ and $n_R$ obtain seesaw masses. Naively, 
if the $SU(2)_R$ breaking scale is $10^{2.5}$ that of $SU(2)_L$, then $m_n \sim 
10^5~m_\nu \sim 10$ keV if $m_\nu \sim 0.1$ eV.  The corresponding Fermi 
constant for $n$ is then $10^{-5}~G_F$, which makes it effectively sterile. 
The lightest scotino $n$ is absolutely stable (having odd $S$) 
in complete analogy to the lightest $\nu$ which is also absolutely stable.  
Note that $n_i \to n_j \gamma$ is possible just 
as $\nu_i \to \nu_j \gamma$ is possible, but these lifetimes are much 
greater than that of the Universe. Consequently, it is possible that 
the three scotinos permeate the Universe as warm dark matter (with 
presumably $\Omega_n h^2 \sim 0.1$), and the three neutrinos as hot 
dark matter (with a much smaller relic density $\Omega_\nu h^2 \sim 0.001$ 
if $\sum m_\nu \sim 0.1$ eV).

The decays of the heavy $\nu_R$ to $\nu \phi_L^0 - e^- \phi_L^+$ and 
$\bar{\nu} \bar{\phi}_L^0 - e^+ \phi_L^-$ and heavy $n_L$ to 
$n \phi_R^0 - e^- \phi_R^+$ and $\bar{n} \bar{\phi}_R^0 - e^+ \phi_R^-$ both 
create lepton asymmetries which get converted to a $B-L$ asymmetry through 
the $SU(2)_L$ and $SU(2)_R$ sphalerons.  Matter and dark matter both have 
$B-L$.  They are distinguished only by $S$.  Once past the electroweak phase 
transition, $n$ becomes effectively a sterile neutrino.  For $m_\nu \sim 
0.1$ eV, its relic density is about $\Omega_\nu h^2 = 0.001$.  To obtain 
the observed dark-matter value $\Omega_n h^2 = 0.1$ for $m_n \sim 10$ keV, 
its number density should be about $10^{-3}$ that of neutrinos. It is 
interesting to note that if $SU(2)_L$ breaks at a much higher scale than 
$SU(2)_R$, then the scotinos may become the neutrinos and vice versa, and the 
Universe would look very much the same, except for a left-right switch.

There are two lepton asymmetries in the early Universe, one each from 
$\nu_R$ and $n_L$ decay.  Their exact values depend on various parameters 
such as the CP violation in the $3 \times 3$ Yukawa coupling matrices in the 
$SU(2)_L$ and $SU(2)_R$ sectors respectively.  As the Universe cools below 
the $SU(2)_R$ 
and $SU(2)_L$ phase transitions, these two asymmetries are converted into 
two $B-L$ asymmetries, one for the $S=0$ fermions, i.e. the usual quarks 
and leptons, and the other for the $S=1$ fermions, i.e. the $h$ quarks 
and the $n$ scotinos.  However, unlike the usual quarks, the $h$ quarks 
all decay away, i.e. $h \to u e \bar{n}$, hence only the $n$ scotinos remain. 
If their present number density is about $10^{-3}$ that of the neutrinos, 
which is presumably a realizable scenario, if for example~\cite{bhl10} a 
late decaying particle exists which releases enough entropy after the 
scotino freeze-out at a temperature of order 1 GeV, then $m_n \sim 10$ 
keV works very well for understanding the structure of the Universe at 
all scales~\cite{dvs11}.  It is especially natural for explaining the 
deficit of dwarf spheroidal galaxies relative to the cold-dark-matter 
scenario.  The role of a late decaying particle may be fulfilled by a 
heavy scotino with a mass of order 1 GeV, as suggested in Ref.~\cite{bhl10}.  
In that case, only two $n$ scotinos are light.  An alternative scenario is to 
have a light scalar singlet which couples to $\Phi^\dagger \Phi$ as in 
many proposed extensions of the standard model invoking the idea of a Higgs 
portal.  The decay of this scalar to photons in one loop may also increase 
the entropy of the Universe by a large factor before big bang 
nucleosynthesis.

As a very weakly interacting particle of 10 keV, $n_R$ is not observable 
at underground experiments searching for dark matter through nuclear recoil. 
Since it does not mix with $\nu_L$, it is also not a product of beta decay.  
It is in fact absolutely stable, so it cannot be directly observed 
astrophysically through radiative decay as is the case with the usual 
sterile neutrino, thus evading the upper bound of about 2 keV from galactic 
X-ray observations.  On the other hand, the lower bound of a few keV still 
holds from Lyman-$\alpha$ forest observations.  As for the one-loop 
annihilation of $nn \to \gamma \gamma$, its cross section is further 
suppressed by the $SU(2)_R$ breaking scale.  This again makes it 
very difficult to observe.  At the Large Hadron 
Collider (LHC), it may show up if there is a light enough $h$ quark, 
say of order 1 TeV.  Such quarks are easily produced in pairs by gluons.  
Once produced, they decay to $n$ and $\bar{n}$, i.e.
\begin{equation}
h \to u \bar{n} e^-, ~~~~~ \bar{h} \to \bar{u} n e^+, 
\end{equation}
through $W_R^\pm$ exchange.  Thus $h$ looks like a fourth-generation $d$ 
quark, which has the current bound of $M > 495$ GeV~\cite{cms11}.  
There is, however, a very important difference which should be explored 
in future LHC analyses.  Whereas a fourth-generation $d$ has suppressed 
branching fractions into leptons, the lightest $h$ decays only 
semileptonically.  This makes it much easier to discover if 
its production is kinematically allowed.  Another possible production 
mechanism is gluon + $u \to h W_R^+$.

This model also predicts two effective Higgs doublets~\cite{bflrss11}, 
i.e. $(\phi_L^+,\phi_L^0)$ and $(\eta_2^+,\eta_2^0)$.  However, unlike the 
canonical case, such as that of the Minimal Supersymmetric Standard Model 
(MSSM), where the $u$ quarks couple to one doublet with 
$m_u \sin^{-1} \beta$, and the $d$ quarks and charged leptons couple to 
the other doublet with $m_d \cos^{-1} \beta$ and $m_e \cos^{-1} 
\beta$, the charged leptons now team up instead with the $u$ quarks 
with coupling $m_e \sin^{-1} \beta$.  This means that 
$H^+ \to \tau^+ \nu_\tau$ is suppressed by $\cot^4 \beta$ relative 
to the prediction of the MSSM.  For large $\tan \beta$, this channel 
would be unobservable and may serve to discriminate 
this two-Higgs-doublet structue against that of the MSSM.

Let $e/g_L = s_L = \sin \theta_W$ and $s_R = e/g_R$ with 
$c_{L,R} = \sqrt{1-s^2_{L.R}}$, then $g_B = e/\sqrt{c_L^2-s_R^2}$ and 
the neutral gauge bosons of this model are given by
\begin{equation}
\pmatrix{A \cr Z \cr Z'} = \pmatrix{s_L & s_R & \sqrt{c_L^2-s_R^2} \cr 
c_L  & -s_L s_R/c_L & -s_L \sqrt{c_L^2-s_R^2}/c_L \cr 0 & \sqrt{c_L^2-s_R^2}/c_L 
& -s_R/c_L} \pmatrix{W_L^0 \cr W_R^0 \cr B}.
\end{equation}
Whereas $Z$ couples to the current $J_Z = J_{3L} - s_L^2 J_{\rm em}$ with 
coupling $e/s_Lc_L$ as in the standard model, $Z'$ couples to the current 
$J_{Z'} = s_R^2 J_{3L} + c_L^2 J_{3R} - s_R^2 J_{\rm em}$ with coupling 
$g_{Z'} = e/s_R c_L \sqrt{c_L^2-s_R^2}$.  Although $Z-Z'$ mixing exists, 
it is negligible because it is of order $M_Z^2/M_{Z'}^2$ which is assumed 
to be $\sim 10^{-5}$.  However, since $Z$ has a component of $W_R^0$, 
it also couples to $W_R^+ W_R^-$.  This appears to induce an effective  
one-loop flavor-changing coupling such as $\bar{e} \gamma^\alpha \mu Z_\alpha$ 
through $n$ exchange or $\bar{u} \gamma^\alpha c Z_\alpha$ through $h$ 
exchange.  A naive calculation of the integral involved seems to indicate 
that this effect is not suppressed if $m_h$ is comparable to $M_{W_R}$, 
which would be a curious example of non-decoupling.  
A full calculation taking into account the contributions of the 
would-be Goldstone modes of the spontaneous symmetry breaking of 
$SU(2)_R$ shows that this effective coulping is in fact zero in the 
limit $SU(2)_L$ is not broken.

If soft breaking of $S'$ is allowed through the terms $\bar{n}_L \nu_R$, 
$\bar{h}_L d_R$, $det(\eta)$, and $\Phi_L^\dagger \tilde{\eta} \Phi_R$, 
then $n$ mixes with $\nu$ (and $h$ with $d$), so the usual sterile-neutrino 
scenario is recovered, but with the addition of the exotic $h$ quarks.

In conclusion, a new left-right extension of the standard model has been 
proposed.  The neutral fermion in the $SU(2)_R$ doublet $(n,e)_R$ is 
not the Dirac mass partner of the neutrino in the $SU(2)_L$ doublet 
$(\nu,e)_L$.  It is in fact a scotino, i.e. an absoultely stable dark-matter 
fermion.  Both $\nu_L$ and $n_R$ obtain seesaw masses from the corresponding 
singlets $\nu_R$ and $n_L$.  Assuming that the $SU(2)_R$ breaking scale is 
$\sim 10^{2.5}$ that of $SU(2)_L$, $m_\nu \sim 0.1$ eV suggests 
$m_n \sim 10$ keV.  It is thus a very suitable candidate for warm dark 
matter, which is a possible 
new paradigm for explaining the structure of the Universe at all scales. 
Matter and dark matter both have baryon and lepton numbers and follow 
parallel paths as the Universe expands.  Both are produced with a $B-L$ 
asymmetry in the early Universe and their remnants are the $u$ and $d$ 
quarks, the electron, three neutrinos as well as three scotinos.

Just as neutrinos are very difficult to see because they interact so 
weakly, the scotinos are much more so.  The only chance may be 
the production of the lightest exotic $h$ quark at the LHC.  Its decay 
is predicted to be completely semileptonic.  This model also predicts 
two Higgs doublets, with one coupling to $d$ quarks, and the other to $u$ 
quarks and charged leptons.  As more LHC data become available, it may 
be possible to test these specific predictions.

This research is supported in part by the U.~S.~Department of Energy 
under Grant No.~DE-AC02-06CH11357.

\bibliographystyle{unsrt}

\end{document}